\documentclass[twocolumn,usenatbib,prl,floats,usegraphicx]{revtex4}

\usepackage{epsfig}
\usepackage{graphicx}
\usepackage{graphics}

\usepackage{amsfonts}  
\usepackage{amsmath, amssymb, stmaryrd, aas_macros}     

\usepackage[usenames]{color}

\newcommand{\fR}{\ensuremath{f(R)~}}
\newcommand{\gC}{\ensuremath{^3G}~}
\newcommand{\gQ}{\ensuremath{^4G}~}

\def\etal{{\frenchspacing\it et al.}}

\def\eg{{\frenchspacing\it e.g.}}

\def\be{\begin{equation}}
\def\ee{\end{equation}}
\def\ba{\begin{eqnarray}}
\def\ea{\end{eqnarray}}

\def\hmpc{h^{-1}\,{\rm Mpc}}

\def\simlt{\stackrel{<}{{}_\sim}}
\def\simgt{\stackrel{>}{{}_\sim}}

\begin{document}

\title{A clear and measurable signature of modified gravity in the galaxy velocity field}

\author{Wojciech A. Hellwing}
\email[Electronic address: ]{wojciech.hellwing@durham.ac.uk}
\affiliation{Institute for Computational Cosmology, Department of Physics, Durham University, South Road, Durham DH1 3LE, UK}
\affiliation{Interdisciplinary Centre for Mathematical and Computational Modelling (ICM), University of Warsaw, ul. Pawi\'nskiego 5a, Warsaw, Poland}
\author{Alexandre Barreira}
\affiliation{Institute for Computational Cosmology, Department of Physics, Durham University, South Road, Durham DH1 3LE, UK}
\affiliation{Institute for Particle Physics Phenomenology,\\ Department of Physics, Durham University, Durham DH1 3LE, U.K.}
\author{Carlos S. Frenk}
\affiliation{Institute for Computational Cosmology, Department of Physics, Durham University, South Road, Durham DH1 3LE, UK}
\author{Baojiu Li}
\affiliation{Institute for Computational Cosmology, Department of Physics, Durham University, South Road, Durham DH1 3LE, UK}
\author{Shaun Cole}
\affiliation{Institute for Computational Cosmology, Department of Physics, Durham University, South Road, Durham DH1 3LE, UK}

\begin{abstract}

  The velocity field of dark matter and galaxies reflects the
  continued action of gravity throughout cosmic history. We show that
  the low-order moments of the pairwise velocity distribution,
  $v_{12}$, are a powerful diagnostic of the laws of gravity on
  cosmological scales. In particular, the projected line-of-sight
  galaxy pairwise velocity dispersion, $\sigma_{12}(r)$, is very
  sensitive to the presence of modified gravity.  Using a set of
  high-resolution N-body simulations we compute the pairwise velocity
  distribution and its projected line-of-sight dispersion for a class
  of modified gravity theories: the chameleon \fR gravity and Galileon
  gravity (cubic and quartic).  The velocities of dark matter halos
  with a wide range of masses would exhibit deviations from General
  Relativity at the $(5-10)\sigma$ level. We examine strategies for
  detecting these deviations in galaxy redshift and peculiar velocity
  surveys. If detected, this signature would be a ``smoking gun'' for
  modified gravity.

\end{abstract}


\maketitle

{\it Introduction.}~ Measurements of temperature anisotropies in the
microwave background radiation and of the large-scale distribution of
galaxies in the local universe have established {\it ``Lambda cold
  dark matter''}, or $\Lambda$CDM, as the standard model of cosmology.
This model is based on Einstein's theory of General Relavity (GR) and
has several parameters that have been determined experimentally to
high precision \citep[\eg][]{2dfgrs,sdss,Hawkins2003,Zehavi2002,bao_2df,bao_sdss,Planck2013}. 
One of these parameters is the
cosmological constant, $\Lambda$, which is responsible for the
accelerating expansion of the Universe but has no known physical basis
within GR. Modifications of GR, generically known as ``modified
gravity'' (MG), could, in principle, provide an explanation (see
e.g.~\citep{Clifton2012} for a comprehensive review).  In this case,
gravity deviates from GR on sufficiently large scales so as to give
rise to the observed accelerated expansion but on small scales such deviations are
suppressed by dynamical screening mechanisms which are
required for these theories to remain compatible with the stringent
tests of gravity in the Solar System \citep{Will:2005va}.

Significant progress has been achieved in recent years in designing
observational tests of gravity on cosmological scales which might
reveal the presence of MG
\citep[\eg][]{Guzzo2008,Weinberg2013,Gbz_prl_2011_detect_mog}. Most
viable MG theories predict changes in the clustering pattern
on non-linear and weakly non-linear scales; on galaxy and halo
dynamics
\citep[\eg][]{Hellwing2013,Barreira2013_cubic,Li2013fofr,rebel1,infall_Zu2013,Jennings2012_rds_fofr,Wojtak2011};
on weak gravitational lensing signals and on the integrated
Sachs-Wolfe effect \citep[\eg][]{halosvoids_fr,fr_Cai2013}.  However,
a common feature of these observational probes is that they typically
rely on quantities for which we have limited model-independent
information due, in part, to various degeneracies, many related to
poorly understood baryonic processes associated with galaxy formation
\citep[\eg][]{bar1,bar2,bar3}. This processes can introduce
further degeneracies in case of MG cosmology \citep{mog_gadget}.
In addition, there are numerous statistical and
systematic uncertainties in the observational data whose size can be 
comparable to the expected deviations from GR.

In this {\it Letter} we introduce the use of the low-order moments of
the distribution of galaxy pairwise velocities as a probe of GR and MG
on cosmological scales. We illustrate the salient physics by reference
to two classes of currently popular MG models. The first is the $f(R)$
family of gravity models \citep{Carrol2004,Chiba2003,SF2010},
in which the Einstein-Hilbert action is
augmented by an arbitrary and intrinsically non-linear function of the
Ricci scalar, $R$. These models include the environment-dependent
``chameleon'' screening mechanism. The second class is {\it Galileon}
gravity \citep{Galileon2009,CovGal2009}, in which the modifications to gravity arise through nonlinear
derivative self-couplings of a Galilean-invariant scalar field.  These
models restore standard gravity on small scales through the Vainshtein
effect \citep{Vainshtein1972}.

Our analysis is based on the high-resolution N-body simulations of
\citep{Li2013fofr}, for the Hu-Sawicki $f(R)$ model \citep{hs2007},
and of \citep{Barreira2013_cubic,Li2013_quartic}, for Galileon gravity
\citep{CovGal2009, Deffayet:2009mn}. These consider three flavours of
$f(R)$ gravity corresponding to different values of the parameter
$|f_{R0}|$ ($10^{-4},10^{-5},10^{-6})$, which determine the degree
of deviation from standard GR \citep{hs2007}.  We refer to these as F4,
F5 and F6 respectively.  For Galileon gravity we study the so-called
Cubic, \gC, and Quartic, \gQ, models, which are characterized by the
order at which the scalar field enters into the Lagrangian
\citep{Galileon2009}.

{\it Pairwise velocities.}~ The mean pairwise relative velocity of
galaxies (or \textit{pairwise streaming velocity}), $v_{12}$, reflects
the ``mean tendency of well-separated galaxies to approach each
other'' \citep{1980Peebles}. This statistic was introduced by
Davis \& Peebels \citep{DavisPeebles_BBGKY} in the context of the
kinetic BBGKY theory \citep{bogoliubov1946,BornGreen1946,Kirkwood1946,yvon1935} 
which describes the dynamical evolution of a
system of particles interacting through gravity.  In the fluid limit
its equivalent is the pair density-weighted relative velocity, \be
\label{eqn:v12-weighted}
\mathbf{v}_{12}(r) = \langle\mathbf{v}_1-\mathbf{v}_2\rangle_{\rho} =
{\langle(\mathbf{v}_1-\mathbf{v}_2)(1+\delta_1)(1+\delta_2)\rangle\over
  1+\xi(r)}\,\,, 
\ee 
where $\mathbf{v}_1$ and $\delta_1=\rho_1/\langle\rho\rangle-1$ denote
the peculiar velocity and fractional matter density contrast 
at position $\mathbf{r}_1$; $r=|\mathbf{r}_1-\mathbf{r}_2|$; and
$\xi(r)=\langle\delta_1\delta_2\rangle$ is the 2-point density
correlation function.  The $\langle\cdots\rangle_{\rho}$ denotes a 
pair-weighted average, which differs from the usual spatial averaging by
the weighting factor, 
$\mathcal{W}=\rho_1\rho_2/\langle\rho_1\rho_2\rangle$. Note that 
$\mathcal{W}$ is proportional to the number density of pairs. 

Gravitational instability theory predicts that the amplitude of
$v_{12}(r)$ is determined by the 2-point correlation function,
$\xi(r)$, and the growth rate of matter density perturbations,
$g\equiv d\ln D_{+}/d\ln a$ (where $D_{+}(a)$ is the linear growing
mode solution and $a$ is the cosmological scale factor) through the
pair conservation equation \citep{1980Peebles}. Juszkiewicz \etal
\citep{Juszkiewicz1999} provided an analytic expression for Eqn.~(\ref{eqn:v12-weighted})
that is a good approximation to the
solution of the pair conservation equation for universes with Gaussian
initial conditions:
$v_{12}=-{2\over3}H_0rg\bar{\bar{\xi}}(r)[1+\alpha\bar{\bar{\xi}}(r)]$,
where
$\bar{\xi}(r)=(3/r^3)\int_0^r\xi(x)x^2dx\equiv\bar{\bar{\xi}}(r)[1+\xi(r)]$.
Here $\alpha$ is a parameter that depends on the logarithmic slope of
$\xi(r)$ and $H_0=100\,h\,$km s$^{-1}$ Mpc$^{-1}$ is the present day
value of the Hubble constant. It is clear that $v_{12}(r)$ is a strong
function of $\xi(r)$ and $g$, both of which will differ in MG theories
from the GR values. This dependency motivates the use of the low-order
moments of the pairwise velocity distribution as tracers of MG and of
the fifth-force it induces on galaxies and dark matter halos. Specifically, we
will consider the following quantities:
\begin{itemize}
 \item the mean radial pairwise velocity, $v_{12}$; 
 \item the dispersion (not centred) of the (radial) pairwise
   velocities, $\sigma_{\parallel}=\langle v_{12}^2\rangle^{1/2}$;  
 \item the mean transverse velocity of pairs, $v_{\perp}$; 
 \item the dispersion of the transverse velocity of pairs,
   $\sigma_{\perp}=\langle v_{\perp}^2\rangle^{1/2}$.
\end{itemize}
Since none of these quantities is directly observable, following
\citep{Jenkins} we also consider the centred line-of-sight 
pairwise velocity dispersion,
$\sigma^2_{12}(r)=\int\xi(R)\sigma^2_{p}(R)dl/\int\xi(R)dl$. Here $r$
is the projected galaxy separation, $R=\sqrt{r^2+l^2}$, and the
integration is taken along the line-of-sight within $l\pm25\hmpc$.  The quantity
$\sigma^2_p$ is the line-of-sight centred pairwise dispersion, defined as in
\citep{Jenkins}:
\be
\label{eqn:sigma_los_p}
\sigma^2_p={r^2\sigma^2_{\perp}/2+l^2(\sigma^2_{\parallel}-v^2_{12})\over r^2+l^2}\,.
\ee

\begin{figure}[!t]
  \begin{center}
    \includegraphics[angle=-90,width=0.48\textwidth]{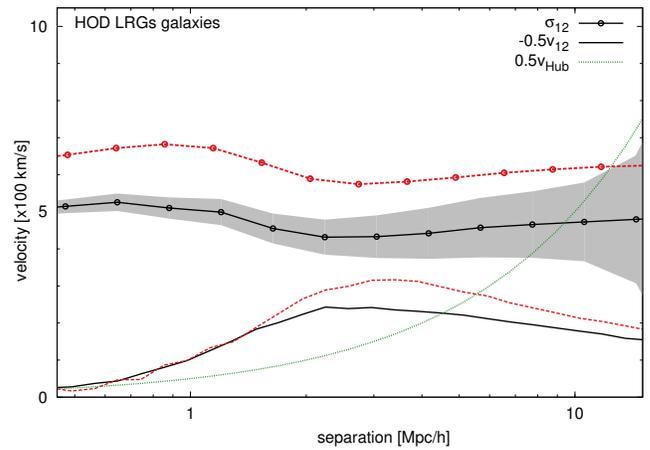}
    \caption{The scale dependence of the pairwise
      velocity moments extracted from HOD mock galaxy catalogues.
      The black solid lines show the GR case, while the red dashed lines show the F4 model. 
      The thin red and black lines show minus the mean streaming 
      velocity, $-v_{12}(r)$, scaled down by factor of 2 for clarity;
      the lines with filled circles show  
      the dispersion, $\sigma_{12}(r)$; The shaded region represents
      an illustrative error as in \citep{Hawkins2003} and \citep{Zehavi2002}.
      The dotted green line shows the Hubble
      velocity, $H_0r$, also scaled down for comparison.  }
    \label{fig:dispersions}
  \end{center}
\end{figure}
\begin{figure*}
  \begin{center}
    \includegraphics[angle=-90,width=0.9\textwidth]{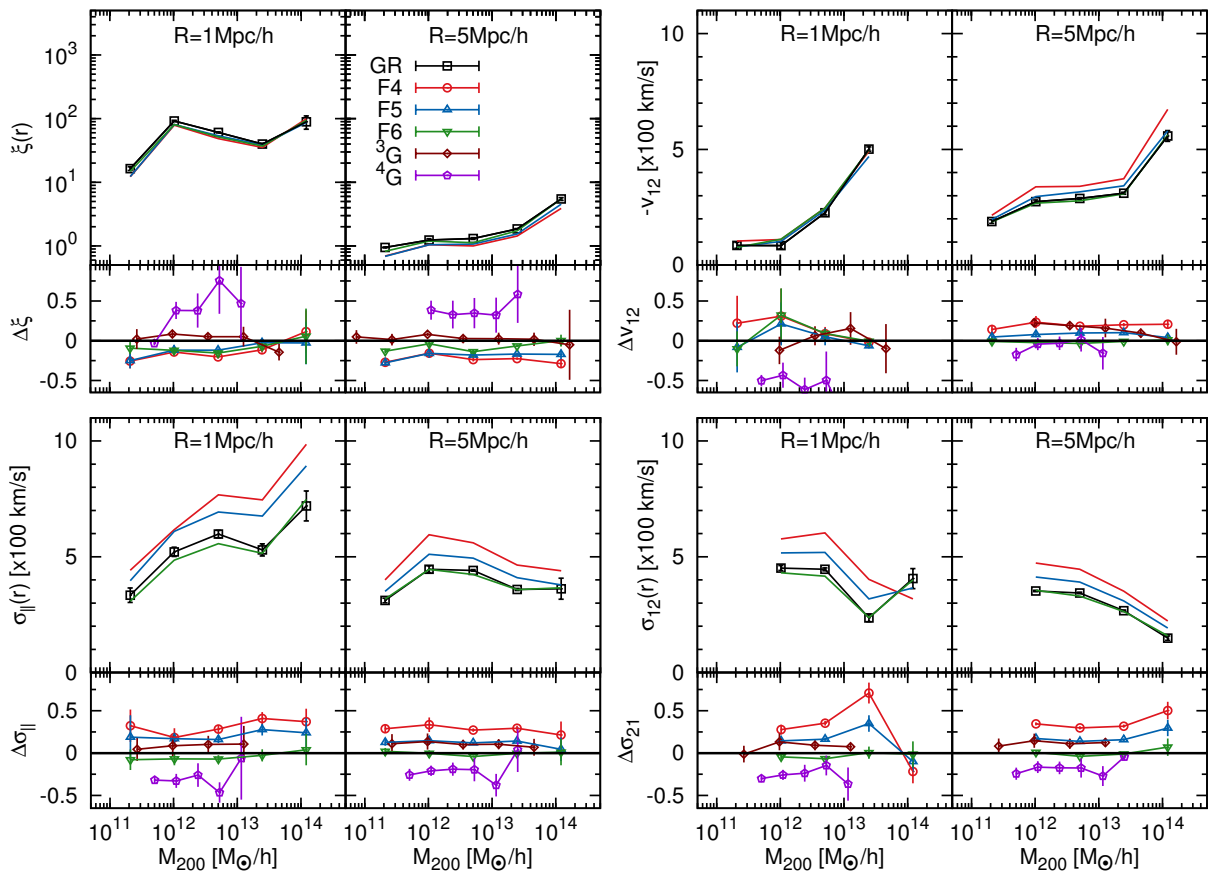}
    \caption{Comparison of absolute values (top panel in each pair)
      and the relative deviation form the GR case 
(bottom panel in each pair) of: the 2-point correlation function, 
$\xi_2(r)$ (top-left panels); minus the mean streaming velocity, 
$-v_{12}(r)$ (top-right panels); the pairwise velocity dispersion,
$\sigma_{\parallel}(r)$ (bottom-left); and the projected pairwise 
velocity dispersion, $\sigma_{12}(r)$. The data are binned in halo
mass, $M_{200}$, and shown at two different pair separations: 
$1$ and $5\hmpc$. The legend in the panel for $\xi_2(5\hmpc)$ gives
the colours and symbols that we use to distinguish the different models.
Top panels show only the LCDM and $f(R)$ cases; the QCDM and Galileons 
were omitted for clarity.}
    \label{fig:mass_binned}
  \end{center}
\end{figure*}

Fig.~\ref{fig:dispersions} shows the scale dependence of the
lower-order moments of the pairwise velocities measured in our N-body
simulations in the GR case (black lines and symbols) and in the F4
model (red lines and symbols).  We choose the F4 model for
illustration because this model is the one for which the chameleon
screening mechanism is the least effective \citep{halosvoids_fr}. 

For the purposes of this comparison, and to allow for a better
connection to observations, we construct mock galaxy catalogues
for these two models by performing a halo occupation distribution (HOD)
analysis \citep[\eg][]{Kravtsov2004}. 
Our HOD catalogues are tuned to resemble a sample of Luminous Red
Galaxies with a satellite fraction of $\sim 7\%$ and a total galaxy
number density of $4 \times 10^{-5} (h/{\rm Mpc})^3$. This number density
is roughly consistent with that of the SDSS DR7 sample presented in \citep{SDSS_LRGs}.
We do this by following
a similar procedure as described in \citep{Wake2008,Barreira_hm_2014}.
The shaded region in the figure shows an illustrative error
that reflects the accuracy of $\sigma_{12}$ measurements form
galaxy redshift surveys as in \citep{Hawkins2003} and \citep{Zehavi2002}.
Firstly, we note that the {\it stable clustering}
regime \citep{1980Peebles} (the scales over which the mean infall
velocity exceeds the Hubble expansion, $-v_{12}>Hr$) extends to larger
separations for the F4 model than for the GR case.  However, $v_{12}$
in F4 differs significantly from GR only in the mildly non-linear regime, 
$2 \simlt r \simlt 10\hmpc$.  The maximum difference between the two
models occurs at $r\sim3.5\hmpc$ and is $\sim 30\%$.  
The situation is quite different when we consider $\sigma_{12}$.
While the F4 values are also roughly $30$ to
$35\%$ larger than in GR, the signal now is noticeable on all
scales plotted. Now, if we compare $\sigma_{12}$ for F4 with the GR case
with errors obtained as in \citep{Hawkins2003,Zehavi2002}, we can see
that the amplitude of this statistics in F4 is $(2-4)\sigma$ away from
the GR case. 

The differences between F4 and GR are driven by the fact that the
distribution of $v_{12}$ never reaches the Gaussian limit, even at
large separations.  This is because, at a given separation, $r$, the
velocity difference between a galaxy pair does not have a net
contribution from modes with wavelengths larger than the pair
separation since those modes make the same contribution to the
velocities of both galaxies.  Hence, on the scale of the typical
interhalo separation (at which the galaxies in a pair inhabit
different halos), the distribution of $v_{12}$ factorises into two
individual peculiar velocity distributions, one for each galaxy or
halo, and these are always sensitive to non-linearities driven by
virial motions within the galaxy host halo (see
\citep{Scoccimarro2004} for more details). In most MG theories
the effects of the fifth force on the dynamics are only significant 
on small nonlinear or mildly nonlinear scales ($\simlt10\hmpc$) which
are probed by the pairwise velocity dispersion. Because of this, the
amplitude of $\sigma_{12}$ is potentially a powerful diagnostic of MG.

The effect of the fifth force on $\sigma_{12}$ is illustrated in
Fig.~\ref{fig:mass_binned} where we plot
$\xi(r)\equiv\langle\delta_1\delta_2\rangle$, $v_{12}$,
$\sigma_{\parallel}$ and $\sigma_{12}$ as a function of 
$M_{200}$ \footnote{$M_{200}\equiv 4/3\pi R^3_{200}200\times\rho_m\Omega_m^{-1}$
is the halo virial mass.} for
the MG models we consider. Results are shown at pair separations
$r=1 \hmpc$ and $5 \hmpc$. Here the error bars show
the variance estimated from the ensemble average of simulations from
different phase realisations of the initial conditions.
We also plot the relative deviation,
$\Delta X=X_{MG}/X_{GR}-1$, from a fiducial model which has the same
expansion history, but includes a fifth force.  This helps identify
changes driven by the modified force law rather than by the modified
expansion dynamics.  For clarity, we only show results for the
Galileon model in the relative difference panels. In the $^4G$ model,
although gravity is enhanced in low-density regions, it is suppressed
in the high-density regions of interest because the Vainshtein
mechanism does not fully screen out all of the modifications to
gravity \citep{Li2013_quartic,Barreira_sph_col_2013}.  This is the
reason why the results for this model point in the opposite direction
to those for the other models (F4, F5, F6 and $^3G$), for which
gravity can only be enhanced by a positive fifth force.  For models
other than $^4G$, Fig.~\ref{fig:mass_binned} shows positive
enhancements relative to GR in $v_{12}, \sigma_{\parallel}$ and
$\sigma_{12}$ but a small reduction in the amplitude of $\xi_2$.
Furthermore, the size of the MG effect in both
$\sigma_{\parallel}$ and $\sigma_{12}$ is approximately independent of
halo mass, although there is a weak trend in $\sigma_{12}$ for the
most massive halos ($M_{200}\simgt 10^{13}M_{\odot}/h$).

The most striking result of this {\it Letter} is the amplitude of the
halo mass-binned $\sigma_{12}$ both at $r=1\hmpc$ and $5\hmpc$.
Relative to GR, the deviations in the F4 model range from $30\%$ to
$75\%$. For the F5 and $^3G$ models, the deviation is smaller, but
still visible at the $\Delta\sigma_{12}\sim0.25$ level. The strong
signal in the amplitude of $\sigma_{12}$ is a combination of the
contributions from $\Delta v_{12}, \Delta \sigma_{\parallel}$ and
$\Delta \sigma_{\perp}$ that are incorporated in $\sigma_p$ as shown
in Eqn~.(\ref{eqn:sigma_los_p}) and from $\Delta\xi_2$ which appears
in the line-of-sight integrals for $\sigma_{12}$. Together, their
combined effect results in a prominent fifth force-like signature.
The amplitude of $\sigma_{12}$ is the strongest observable deviation
from GR on cosmological scales so far identified, a potential {\it 
  smoking gun} for MG. This signal, however, is not entirely generic.
For example, the F6 model is virtually indistinguishable from GR: the
fifth force in this flavour of \fR gravity is much too weak to produce
a detectable effect in the dynamics of galaxies and halos.

{\it Summary.}~ Using dark matter halo catalogues extracted from
high-resolution N-body simulations of the formation of cosmic
structure in two representative classes of modified gravity theories
we have computed the mean pairwise streaming velocity and its
dispersion (radial and projected along the line-of-sight). Our
simulations show that there is a strong MG signal contained in
the line-of-sight projected pairwise velocity dispersion. For the
F5,\gC and \gQ models, deviations from GR are at the $>5\sigma$ level
for all masses. The deviation is even more pronounced for the F4
model, where it is at the $>10\sigma$ level and higher.  This is the
clearest footprint of modified gravity found to date in quantities
that are, in principle, observable. Nonetheless, in
a realistic observational situation one can expect the significance 
of the MG signal to be reduced due to ambiguities related to galaxy formation
and observational errors, as illustrated by our HOD
analysis. However, the quality of the data as used by \citep{Hawkins2003,Zehavi2002}
would already be enough to distinguish between GR and
F4, F5 and \gC at the $2\sigma$ level, and these are relatively
older datasets. With current and future surveys like SDSS-II, BOSS
and Pan-STARRS1 \citep[\eg][]{SDSSDR8,BOSS1,BOSS2,PANSTARRS1,PANSTARRS2} 
one can hope to do better, since the new data provide already $\sim30\%$ improved accuracy.

The remaining important question is whether the MG footprint we
have identified is actually observable in the real Universe. As
mentioned above, the $\sigma_{12}(r)$ value can be estimated from
galaxy redshift survey data but only in a model-dependent way.
Specifically, one can obtain the line-of-sight dispersion by fitting
the 2D galaxy redshift space correlation function to a model,
$\xi^s(r_p,\pi)=\int\xi^{'}(r_p,\pi-v/H_0)h(v_{12})\textrm{d}v$, where
$\xi^{'}$ is the linear theory model prediction (which depends on
coherent infall velocities) and the convolution is made with the
assumed distribution of pairwise velocities, $h(v_{12})$
\citep{1980Peebles,1983Davis_Peebles,Benson2000_pvd_gal,Scoccimarro2004}.  
Alternatively, one can use the redshift space
power spectrum of the galaxy distribution to derive a quantity in
Fourier space, $\sigma_{12}(k)$, which is not an exact equivalent of
the configuration space dispersion, but is closely related to it
\citep[\eg][]{Li2006_pvd_gal,Tinker2007_pvd_gal,delaTorre_rsd}. To
apply either of these methods one needs a self-consistent model of the
redshift-space clustering expected in a given MG theory. In
particular, such a model needs to describe the linear galaxy bias
parameter, $b$; the linear growth rate of matter, $g$; and the
pairwise velocity distribution in configuration space, $h(v_{12})$,
or, equivalently, the damping function in Fourier space,
$D[k\mu\sigma_{12}(k)]$. Fortunately, all these quantities can be
derived self-consistently for MG theories using linear
perturbation theory complemented with N-body simulations. Such a
programme is currently being developed.

Instead of using redshift data, it is possible, in principle, to
estimate $v_{12}$ and $\sigma_{12}$ directly from measurements of
galaxy peculiar velocities. The advantage of this approach is that it
is model independent. The disadvantage is that peculiar velocities can
only be measured with sufficient accuracy for a small sample of local
galaxies ($z<0.05$) and even then there are potentially large
systematic errors in the estimates of redshift-independent distance
indicators \citep[\eg][]{StraussWillick,Tully2013}.  A further
complication is that only the radial component of a galaxy peculiar
velocity is observable (but see \citep{Nusser2012}), so it is
necessary to construct special estimators for pairwise velocities such
as those proposed by
\citep{Gorski1989,Juszkiewicz2000,streaming_vel_Omega,Ferreira1999}.

There is already a large body of velocity data of potentially
sufficient quality for the test we propose (\textit{cf.\ } the size of
the velocity error bars in Fig.~23 of \citep{Hawkins2003}). Further
theoretical work is required to refine the redshift-space probes and
further observational work to exploit direct peculiar velocity
measurements. It is to be hoped that the presence of a fifth force, if
it exists, will be revealed in measurements of the galaxy velocity
field.

\bibliography{smoking_gun}

\begin{thebibliography}{65}
\expandafter\ifx\csname natexlab\endcsname\relax\def\natexlab#1{#1}\fi
\expandafter\ifx\csname bibnamefont\endcsname\relax
  \def\bibnamefont#1{#1}\fi
\expandafter\ifx\csname bibfnamefont\endcsname\relax
  \def\bibfnamefont#1{#1}\fi
\expandafter\ifx\csname citenamefont\endcsname\relax
  \def\citenamefont#1{#1}\fi
\expandafter\ifx\csname url\endcsname\relax
  \def\url#1{\texttt{#1}}\fi
\expandafter\ifx\csname urlprefix\endcsname\relax\def\urlprefix{URL }\fi
\providecommand{\bibinfo}[2]{#2}
\providecommand{\eprint}[2][]{\url{#2}}

\bibitem[{\citenamefont{{Colless} et~al.}(2001)\citenamefont{{Colless},
  {Dalton}, {Maddox}, {Sutherland}, {Norberg}, {Cole}, {Bland-Hawthorn},
  {Bridges}, {Cannon}, {Collins} et~al.}}]{2dfgrs}
\bibinfo{author}{\bibfnamefont{M.}~\bibnamefont{{Colless}}},
  \bibinfo{author}{\bibfnamefont{G.}~\bibnamefont{{Dalton}}},
  \bibinfo{author}{\bibfnamefont{S.}~\bibnamefont{{Maddox}}},
  \bibinfo{author}{\bibfnamefont{W.}~\bibnamefont{{Sutherland}}},
  \bibinfo{author}{\bibfnamefont{P.}~\bibnamefont{{Norberg}}},
  \bibinfo{author}{\bibfnamefont{S.}~\bibnamefont{{Cole}}},
  \bibinfo{author}{\bibfnamefont{J.}~\bibnamefont{{Bland-Hawthorn}}},
  \bibinfo{author}{\bibfnamefont{T.}~\bibnamefont{{Bridges}}},
  \bibinfo{author}{\bibfnamefont{R.}~\bibnamefont{{Cannon}}},
  \bibinfo{author}{\bibfnamefont{C.}~\bibnamefont{{Collins}}},
  \bibnamefont{et~al.}, \bibinfo{journal}{\mnras}
  \textbf{\bibinfo{volume}{328}}, \bibinfo{pages}{1039} (\bibinfo{year}{2001}),
  \eprint{astro-ph/0106498}.

\bibitem[{\citenamefont{{Tegmark} et~al.}(2004)\citenamefont{{Tegmark},
  {Strauss}, {Blanton}, {Abazajian}, {Dodelson}, {Sandvik}, {Wang}, {Weinberg},
  {Zehavi}, {Bahcall} et~al.}}]{sdss}
\bibinfo{author}{\bibfnamefont{M.}~\bibnamefont{{Tegmark}}},
  \bibinfo{author}{\bibfnamefont{M.~A.} \bibnamefont{{Strauss}}},
  \bibinfo{author}{\bibfnamefont{M.~R.} \bibnamefont{{Blanton}}},
  \bibinfo{author}{\bibfnamefont{K.}~\bibnamefont{{Abazajian}}},
  \bibinfo{author}{\bibfnamefont{S.}~\bibnamefont{{Dodelson}}},
  \bibinfo{author}{\bibfnamefont{H.}~\bibnamefont{{Sandvik}}},
  \bibinfo{author}{\bibfnamefont{X.}~\bibnamefont{{Wang}}},
  \bibinfo{author}{\bibfnamefont{D.~H.} \bibnamefont{{Weinberg}}},
  \bibinfo{author}{\bibfnamefont{I.}~\bibnamefont{{Zehavi}}},
  \bibinfo{author}{\bibfnamefont{N.~A.} \bibnamefont{{Bahcall}}},
  \bibnamefont{et~al.}, \bibinfo{journal}{\prd} \textbf{\bibinfo{volume}{69}},
  \bibinfo{eid}{103501} (\bibinfo{year}{2004}), \eprint{astro-ph/0310723}.

\bibitem[{\citenamefont{{Hawkins} et~al.}(2003)\citenamefont{{Hawkins},
  {Maddox}, {Cole}, {Lahav}, {Madgwick}, {Norberg}, {Peacock}, {Baldry},
  {Baugh}, {Bland-Hawthorn} et~al.}}]{Hawkins2003}
\bibinfo{author}{\bibfnamefont{E.}~\bibnamefont{{Hawkins}}},
  \bibinfo{author}{\bibfnamefont{S.}~\bibnamefont{{Maddox}}},
  \bibinfo{author}{\bibfnamefont{S.}~\bibnamefont{{Cole}}},
  \bibinfo{author}{\bibfnamefont{O.}~\bibnamefont{{Lahav}}},
  \bibinfo{author}{\bibfnamefont{D.~S.} \bibnamefont{{Madgwick}}},
  \bibinfo{author}{\bibfnamefont{P.}~\bibnamefont{{Norberg}}},
  \bibinfo{author}{\bibfnamefont{J.~A.} \bibnamefont{{Peacock}}},
  \bibinfo{author}{\bibfnamefont{I.~K.} \bibnamefont{{Baldry}}},
  \bibinfo{author}{\bibfnamefont{C.~M.} \bibnamefont{{Baugh}}},
  \bibinfo{author}{\bibfnamefont{J.}~\bibnamefont{{Bland-Hawthorn}}},
  \bibnamefont{et~al.}, \bibinfo{journal}{\mnras}
  \textbf{\bibinfo{volume}{346}}, \bibinfo{pages}{78} (\bibinfo{year}{2003}),
  \eprint{astro-ph/0212375}.

\bibitem[{\citenamefont{{Zehavi} et~al.}(2002)\citenamefont{{Zehavi},
  {Blanton}, {Frieman}, {Weinberg}, {Mo}, {Strauss}, {Anderson}, {Annis},
  {Bahcall}, {Bernardi} et~al.}}]{Zehavi2002}
\bibinfo{author}{\bibfnamefont{I.}~\bibnamefont{{Zehavi}}},
  \bibinfo{author}{\bibfnamefont{M.~R.} \bibnamefont{{Blanton}}},
  \bibinfo{author}{\bibfnamefont{J.~A.} \bibnamefont{{Frieman}}},
  \bibinfo{author}{\bibfnamefont{D.~H.} \bibnamefont{{Weinberg}}},
  \bibinfo{author}{\bibfnamefont{H.~J.} \bibnamefont{{Mo}}},
  \bibinfo{author}{\bibfnamefont{M.~A.} \bibnamefont{{Strauss}}},
  \bibinfo{author}{\bibfnamefont{S.~F.} \bibnamefont{{Anderson}}},
  \bibinfo{author}{\bibfnamefont{J.}~\bibnamefont{{Annis}}},
  \bibinfo{author}{\bibfnamefont{N.~A.} \bibnamefont{{Bahcall}}},
  \bibinfo{author}{\bibfnamefont{M.}~\bibnamefont{{Bernardi}}},
  \bibnamefont{et~al.}, \bibinfo{journal}{\apj} \textbf{\bibinfo{volume}{571}},
  \bibinfo{pages}{172} (\bibinfo{year}{2002}), \eprint{astro-ph/0106476}.

\bibitem[{\citenamefont{{Cole} et~al.}(2005)\citenamefont{{Cole}, {Percival},
  {Peacock}, {Norberg}, {Baugh}, {Frenk}, {Baldry}, {Bland-Hawthorn},
  {Bridges}, {Cannon} et~al.}}]{bao_2df}
\bibinfo{author}{\bibfnamefont{S.}~\bibnamefont{{Cole}}},
  \bibinfo{author}{\bibfnamefont{W.~J.} \bibnamefont{{Percival}}},
  \bibinfo{author}{\bibfnamefont{J.~A.} \bibnamefont{{Peacock}}},
  \bibinfo{author}{\bibfnamefont{P.}~\bibnamefont{{Norberg}}},
  \bibinfo{author}{\bibfnamefont{C.~M.} \bibnamefont{{Baugh}}},
  \bibinfo{author}{\bibfnamefont{C.~S.} \bibnamefont{{Frenk}}},
  \bibinfo{author}{\bibfnamefont{I.}~\bibnamefont{{Baldry}}},
  \bibinfo{author}{\bibfnamefont{J.}~\bibnamefont{{Bland-Hawthorn}}},
  \bibinfo{author}{\bibfnamefont{T.}~\bibnamefont{{Bridges}}},
  \bibinfo{author}{\bibfnamefont{R.}~\bibnamefont{{Cannon}}},
  \bibnamefont{et~al.}, \bibinfo{journal}{\mnras}
  \textbf{\bibinfo{volume}{362}}, \bibinfo{pages}{505} (\bibinfo{year}{2005}),
  \eprint{astro-ph/0501174}.

\bibitem[{\citenamefont{{Eisenstein} et~al.}(2005)\citenamefont{{Eisenstein},
  {Zehavi}, {Hogg}, {Scoccimarro}, {Blanton}, {Nichol}, {Scranton}, {Seo},
  {Tegmark}, {Zheng} et~al.}}]{bao_sdss}
\bibinfo{author}{\bibfnamefont{D.~J.} \bibnamefont{{Eisenstein}}},
  \bibinfo{author}{\bibfnamefont{I.}~\bibnamefont{{Zehavi}}},
  \bibinfo{author}{\bibfnamefont{D.~W.} \bibnamefont{{Hogg}}},
  \bibinfo{author}{\bibfnamefont{R.}~\bibnamefont{{Scoccimarro}}},
  \bibinfo{author}{\bibfnamefont{M.~R.} \bibnamefont{{Blanton}}},
  \bibinfo{author}{\bibfnamefont{R.~C.} \bibnamefont{{Nichol}}},
  \bibinfo{author}{\bibfnamefont{R.}~\bibnamefont{{Scranton}}},
  \bibinfo{author}{\bibfnamefont{H.-J.} \bibnamefont{{Seo}}},
  \bibinfo{author}{\bibfnamefont{M.}~\bibnamefont{{Tegmark}}},
  \bibinfo{author}{\bibfnamefont{Z.}~\bibnamefont{{Zheng}}},
  \bibnamefont{et~al.}, \bibinfo{journal}{\apj} \textbf{\bibinfo{volume}{633}},
  \bibinfo{pages}{560} (\bibinfo{year}{2005}), \eprint{astro-ph/0501171}.

\bibitem[{\citenamefont{{The Planck Collaboration}
  et~al.}(2013)\citenamefont{{The Planck Collaboration}, {Ade}, {Aghanim},
  {Armitage-Caplan}, {Arnaud}, {Ashdown}, {Atrio-Barandela}, {Aumont},
  {Baccigalupi}, {Banday} et~al.}}]{Planck2013}
\bibinfo{author}{\bibnamefont{{The Planck Collaboration}}},
  \bibinfo{author}{\bibfnamefont{P.~A.~R.} \bibnamefont{{Ade}}},
  \bibinfo{author}{\bibfnamefont{N.}~\bibnamefont{{Aghanim}}},
  \bibinfo{author}{\bibfnamefont{C.}~\bibnamefont{{Armitage-Caplan}}},
  \bibinfo{author}{\bibfnamefont{M.}~\bibnamefont{{Arnaud}}},
  \bibinfo{author}{\bibfnamefont{M.}~\bibnamefont{{Ashdown}}},
  \bibinfo{author}{\bibfnamefont{F.}~\bibnamefont{{Atrio-Barandela}}},
  \bibinfo{author}{\bibfnamefont{J.}~\bibnamefont{{Aumont}}},
  \bibinfo{author}{\bibfnamefont{C.}~\bibnamefont{{Baccigalupi}}},
  \bibinfo{author}{\bibfnamefont{A.~J.} \bibnamefont{{Banday}}},
  \bibnamefont{et~al.}, \bibinfo{journal}{ArXiv e-prints}
  (\bibinfo{year}{2013}), \eprint{1303.5076}.

\bibitem[{\citenamefont{{Clifton} et~al.}(2012)\citenamefont{{Clifton},
  {Ferreira}, {Padilla}, and {Skordis}}}]{Clifton2012}
\bibinfo{author}{\bibfnamefont{T.}~\bibnamefont{{Clifton}}},
  \bibinfo{author}{\bibfnamefont{P.~G.} \bibnamefont{{Ferreira}}},
  \bibinfo{author}{\bibfnamefont{A.}~\bibnamefont{{Padilla}}},
  \bibnamefont{and}
  \bibinfo{author}{\bibfnamefont{C.}~\bibnamefont{{Skordis}}},
  \bibinfo{journal}{\physrep} \textbf{\bibinfo{volume}{513}},
  \bibinfo{pages}{1} (\bibinfo{year}{2012}), \eprint{1106.2476}.

\bibitem[{\citenamefont{{Will}}(2006)}]{Will:2005va}
\bibinfo{author}{\bibfnamefont{C.~M.} \bibnamefont{{Will}}},
  \bibinfo{journal}{Living Reviews in Relativity} \textbf{\bibinfo{volume}{9}},
  \bibinfo{pages}{3} (\bibinfo{year}{2006}), \eprint{gr-qc/0510072}.

\bibitem[{\citenamefont{{Guzzo} et~al.}(2008)\citenamefont{{Guzzo},
  {Pierleoni}, {Meneux}, {Branchini}, {Le F{\`e}vre}, {Marinoni}, {Garilli},
  {Blaizot}, {De Lucia}, {Pollo} et~al.}}]{Guzzo2008}
\bibinfo{author}{\bibfnamefont{L.}~\bibnamefont{{Guzzo}}},
  \bibinfo{author}{\bibfnamefont{M.}~\bibnamefont{{Pierleoni}}},
  \bibinfo{author}{\bibfnamefont{B.}~\bibnamefont{{Meneux}}},
  \bibinfo{author}{\bibfnamefont{E.}~\bibnamefont{{Branchini}}},
  \bibinfo{author}{\bibfnamefont{O.}~\bibnamefont{{Le F{\`e}vre}}},
  \bibinfo{author}{\bibfnamefont{C.}~\bibnamefont{{Marinoni}}},
  \bibinfo{author}{\bibfnamefont{B.}~\bibnamefont{{Garilli}}},
  \bibinfo{author}{\bibfnamefont{J.}~\bibnamefont{{Blaizot}}},
  \bibinfo{author}{\bibfnamefont{G.}~\bibnamefont{{De Lucia}}},
  \bibinfo{author}{\bibfnamefont{A.}~\bibnamefont{{Pollo}}},
  \bibnamefont{et~al.}, \bibinfo{journal}{\nat} \textbf{\bibinfo{volume}{451}},
  \bibinfo{pages}{541} (\bibinfo{year}{2008}), \eprint{0802.1944}.

\bibitem[{\citenamefont{{Weinberg} et~al.}(2013)\citenamefont{{Weinberg},
  {Mortonson}, {Eisenstein}, {Hirata}, {Riess}, and {Rozo}}}]{Weinberg2013}
\bibinfo{author}{\bibfnamefont{D.~H.} \bibnamefont{{Weinberg}}},
  \bibinfo{author}{\bibfnamefont{M.~J.} \bibnamefont{{Mortonson}}},
  \bibinfo{author}{\bibfnamefont{D.~J.} \bibnamefont{{Eisenstein}}},
  \bibinfo{author}{\bibfnamefont{C.}~\bibnamefont{{Hirata}}},
  \bibinfo{author}{\bibfnamefont{A.~G.} \bibnamefont{{Riess}}},
  \bibnamefont{and} \bibinfo{author}{\bibfnamefont{E.}~\bibnamefont{{Rozo}}},
  \bibinfo{journal}{\physrep} \textbf{\bibinfo{volume}{530}},
  \bibinfo{pages}{87} (\bibinfo{year}{2013}), \eprint{1201.2434}.

\bibitem[{\citenamefont{{Zhao} et~al.}(2011)\citenamefont{{Zhao}, {Li}, and
  {Koyama}}}]{Gbz_prl_2011_detect_mog}
\bibinfo{author}{\bibfnamefont{G.-B.} \bibnamefont{{Zhao}}},
  \bibinfo{author}{\bibfnamefont{B.}~\bibnamefont{{Li}}}, \bibnamefont{and}
  \bibinfo{author}{\bibfnamefont{K.}~\bibnamefont{{Koyama}}},
  \bibinfo{journal}{Physical Review Letters} \textbf{\bibinfo{volume}{107}},
  \bibinfo{eid}{071303} (\bibinfo{year}{2011}), \eprint{1105.0922}.

\bibitem[{\citenamefont{{Hellwing} et~al.}(2013)\citenamefont{{Hellwing}, {Li},
  {Frenk}, and {Cole}}}]{Hellwing2013}
\bibinfo{author}{\bibfnamefont{W.~A.} \bibnamefont{{Hellwing}}},
  \bibinfo{author}{\bibfnamefont{B.}~\bibnamefont{{Li}}},
  \bibinfo{author}{\bibfnamefont{C.~S.} \bibnamefont{{Frenk}}},
  \bibnamefont{and} \bibinfo{author}{\bibfnamefont{S.}~\bibnamefont{{Cole}}},
  \bibinfo{journal}{\mnras} \textbf{\bibinfo{volume}{435}},
  \bibinfo{pages}{2806} (\bibinfo{year}{2013}), \eprint{1305.7486}.

\bibitem[{\citenamefont{{Barreira}
  et~al.}(2013{\natexlab{a}})\citenamefont{{Barreira}, {Li}, {Hellwing},
  {Baugh}, and {Pascoli}}}]{Barreira2013_cubic}
\bibinfo{author}{\bibfnamefont{A.}~\bibnamefont{{Barreira}}},
  \bibinfo{author}{\bibfnamefont{B.}~\bibnamefont{{Li}}},
  \bibinfo{author}{\bibfnamefont{W.~A.} \bibnamefont{{Hellwing}}},
  \bibinfo{author}{\bibfnamefont{C.~M.} \bibnamefont{{Baugh}}},
  \bibnamefont{and}
  \bibinfo{author}{\bibfnamefont{S.}~\bibnamefont{{Pascoli}}},
  \bibinfo{journal}{\jcap} \textbf{\bibinfo{volume}{10}}, \bibinfo{eid}{027}
  (\bibinfo{year}{2013}{\natexlab{a}}), \eprint{1306.3219}.

\bibitem[{\citenamefont{{Li} et~al.}(2013{\natexlab{a}})\citenamefont{{Li},
  {Hellwing}, {Koyama}, {Zhao}, {Jennings}, and {Baugh}}}]{Li2013fofr}
\bibinfo{author}{\bibfnamefont{B.}~\bibnamefont{{Li}}},
  \bibinfo{author}{\bibfnamefont{W.~A.} \bibnamefont{{Hellwing}}},
  \bibinfo{author}{\bibfnamefont{K.}~\bibnamefont{{Koyama}}},
  \bibinfo{author}{\bibfnamefont{G.-B.} \bibnamefont{{Zhao}}},
  \bibinfo{author}{\bibfnamefont{E.}~\bibnamefont{{Jennings}}},
  \bibnamefont{and} \bibinfo{author}{\bibfnamefont{C.~M.}
  \bibnamefont{{Baugh}}}, \bibinfo{journal}{\mnras}
  \textbf{\bibinfo{volume}{428}}, \bibinfo{pages}{743}
  (\bibinfo{year}{2013}{\natexlab{a}}), \eprint{1206.4317}.

\bibitem[{\citenamefont{{Hellwing} and {Juszkiewicz}}(2009)}]{rebel1}
\bibinfo{author}{\bibfnamefont{W.~A.} \bibnamefont{{Hellwing}}}
  \bibnamefont{and}
  \bibinfo{author}{\bibfnamefont{R.}~\bibnamefont{{Juszkiewicz}}},
  \bibinfo{journal}{Phys.~Rev.~D} \textbf{\bibinfo{volume}{80}},
  \bibinfo{pages}{083522} (\bibinfo{year}{2009}), \eprint{0809.1976}.

\bibitem[{\citenamefont{{Zu} et~al.}(2013)\citenamefont{{Zu}, {Weinberg},
  {Jennings}, {Li}, and {Wyman}}}]{infall_Zu2013}
\bibinfo{author}{\bibfnamefont{Y.}~\bibnamefont{{Zu}}},
  \bibinfo{author}{\bibfnamefont{D.~H.} \bibnamefont{{Weinberg}}},
  \bibinfo{author}{\bibfnamefont{E.}~\bibnamefont{{Jennings}}},
  \bibinfo{author}{\bibfnamefont{B.}~\bibnamefont{{Li}}}, \bibnamefont{and}
  \bibinfo{author}{\bibfnamefont{M.}~\bibnamefont{{Wyman}}},
  \bibinfo{journal}{ArXiv e-prints}  (\bibinfo{year}{2013}),
  \eprint{1310.6768}.

\bibitem[{\citenamefont{{Jennings} et~al.}(2012)\citenamefont{{Jennings},
  {Baugh}, {Li}, {Zhao}, and {Koyama}}}]{Jennings2012_rds_fofr}
\bibinfo{author}{\bibfnamefont{E.}~\bibnamefont{{Jennings}}},
  \bibinfo{author}{\bibfnamefont{C.~M.} \bibnamefont{{Baugh}}},
  \bibinfo{author}{\bibfnamefont{B.}~\bibnamefont{{Li}}},
  \bibinfo{author}{\bibfnamefont{G.-B.} \bibnamefont{{Zhao}}},
  \bibnamefont{and} \bibinfo{author}{\bibfnamefont{K.}~\bibnamefont{{Koyama}}},
  \bibinfo{journal}{\mnras} \textbf{\bibinfo{volume}{425}},
  \bibinfo{pages}{2128} (\bibinfo{year}{2012}), \eprint{1205.2698}.

\bibitem[{\citenamefont{{Wojtak} et~al.}(2011)\citenamefont{{Wojtak}, {Hansen},
  and {Hjorth}}}]{Wojtak2011}
\bibinfo{author}{\bibfnamefont{R.}~\bibnamefont{{Wojtak}}},
  \bibinfo{author}{\bibfnamefont{S.~H.} \bibnamefont{{Hansen}}},
  \bibnamefont{and} \bibinfo{author}{\bibfnamefont{J.}~\bibnamefont{{Hjorth}}},
  \bibinfo{journal}{\nat} \textbf{\bibinfo{volume}{477}}, \bibinfo{pages}{567}
  (\bibinfo{year}{2011}), \eprint{1109.6571}.

\bibitem[{\citenamefont{{Li} et~al.}(2012)\citenamefont{{Li}, {Zhao}, and
  {Koyama}}}]{halosvoids_fr}
\bibinfo{author}{\bibfnamefont{B.}~\bibnamefont{{Li}}},
  \bibinfo{author}{\bibfnamefont{G.-B.} \bibnamefont{{Zhao}}},
  \bibnamefont{and} \bibinfo{author}{\bibfnamefont{K.}~\bibnamefont{{Koyama}}},
  \bibinfo{journal}{\mnras} \textbf{\bibinfo{volume}{421}},
  \bibinfo{pages}{3481} (\bibinfo{year}{2012}), \eprint{1111.2602}.

\bibitem[{\citenamefont{{Cai} et~al.}(2013)\citenamefont{{Cai}, {Li}, {Cole},
  {Frenk}, and {Neyrinck}}}]{fr_Cai2013}
\bibinfo{author}{\bibfnamefont{Y.-C.} \bibnamefont{{Cai}}},
  \bibinfo{author}{\bibfnamefont{B.}~\bibnamefont{{Li}}},
  \bibinfo{author}{\bibfnamefont{S.}~\bibnamefont{{Cole}}},
  \bibinfo{author}{\bibfnamefont{C.~S.} \bibnamefont{{Frenk}}},
  \bibnamefont{and}
  \bibinfo{author}{\bibfnamefont{M.}~\bibnamefont{{Neyrinck}}},
  \bibinfo{journal}{ArXiv e-prints}  (\bibinfo{year}{2013}),
  \eprint{1310.6986}.

\bibitem[{\citenamefont{{White} and {Rees}}(1978)}]{bar1}
\bibinfo{author}{\bibfnamefont{S.~D.~M.} \bibnamefont{{White}}}
  \bibnamefont{and} \bibinfo{author}{\bibfnamefont{M.~J.}
  \bibnamefont{{Rees}}}, \bibinfo{journal}{\mnras}
  \textbf{\bibinfo{volume}{183}}, \bibinfo{pages}{341} (\bibinfo{year}{1978}).

\bibitem[{\citenamefont{{Croton} et~al.}(2006)\citenamefont{{Croton},
  {Springel}, {White}, {De Lucia}, {Frenk}, {Gao}, {Jenkins}, {Kauffmann},
  {Navarro}, and {Yoshida}}}]{bar2}
\bibinfo{author}{\bibfnamefont{D.~J.} \bibnamefont{{Croton}}},
  \bibinfo{author}{\bibfnamefont{V.}~\bibnamefont{{Springel}}},
  \bibinfo{author}{\bibfnamefont{S.~D.~M.} \bibnamefont{{White}}},
  \bibinfo{author}{\bibfnamefont{G.}~\bibnamefont{{De Lucia}}},
  \bibinfo{author}{\bibfnamefont{C.~S.} \bibnamefont{{Frenk}}},
  \bibinfo{author}{\bibfnamefont{L.}~\bibnamefont{{Gao}}},
  \bibinfo{author}{\bibfnamefont{A.}~\bibnamefont{{Jenkins}}},
  \bibinfo{author}{\bibfnamefont{G.}~\bibnamefont{{Kauffmann}}},
  \bibinfo{author}{\bibfnamefont{J.~F.} \bibnamefont{{Navarro}}},
  \bibnamefont{and}
  \bibinfo{author}{\bibfnamefont{N.}~\bibnamefont{{Yoshida}}},
  \bibinfo{journal}{\mnras} \textbf{\bibinfo{volume}{365}}, \bibinfo{pages}{11}
  (\bibinfo{year}{2006}), \eprint{arXiv:astro-ph/0508046}.

\bibitem[{\citenamefont{{White} and {Frenk}}(1991)}]{bar3}
\bibinfo{author}{\bibfnamefont{S.~D.~M.} \bibnamefont{{White}}}
  \bibnamefont{and} \bibinfo{author}{\bibfnamefont{C.~S.}
  \bibnamefont{{Frenk}}}, \bibinfo{journal}{\apj}
  \textbf{\bibinfo{volume}{379}}, \bibinfo{pages}{52} (\bibinfo{year}{1991}).

\bibitem[{\citenamefont{{Puchwein} et~al.}(2013)\citenamefont{{Puchwein},
  {Baldi}, and {Springel}}}]{mog_gadget}
\bibinfo{author}{\bibfnamefont{E.}~\bibnamefont{{Puchwein}}},
  \bibinfo{author}{\bibfnamefont{M.}~\bibnamefont{{Baldi}}}, \bibnamefont{and}
  \bibinfo{author}{\bibfnamefont{V.}~\bibnamefont{{Springel}}},
  \bibinfo{journal}{ArXiv e-prints}  (\bibinfo{year}{2013}),
  \eprint{1305.2418}.

\bibitem[{\citenamefont{{Carroll} et~al.}(2004)\citenamefont{{Carroll},
  {Duvvuri}, {Trodden}, and {Turner}}}]{Carrol2004}
\bibinfo{author}{\bibfnamefont{S.~M.} \bibnamefont{{Carroll}}},
  \bibinfo{author}{\bibfnamefont{V.}~\bibnamefont{{Duvvuri}}},
  \bibinfo{author}{\bibfnamefont{M.}~\bibnamefont{{Trodden}}},
  \bibnamefont{and} \bibinfo{author}{\bibfnamefont{M.~S.}
  \bibnamefont{{Turner}}}, \bibinfo{journal}{\prd}
  \textbf{\bibinfo{volume}{70}}, \bibinfo{eid}{043528} (\bibinfo{year}{2004}),
  \eprint{astro-ph/0306438}.

\bibitem[{\citenamefont{{Chiba}}(2003)}]{Chiba2003}
\bibinfo{author}{\bibfnamefont{T.}~\bibnamefont{{Chiba}}},
  \bibinfo{journal}{Physics Letters B} \textbf{\bibinfo{volume}{575}},
  \bibinfo{pages}{1} (\bibinfo{year}{2003}), \eprint{astro-ph/0307338}.

\bibitem[{\citenamefont{{Sotiriou} and {Faraoni}}(2010)}]{SF2010}
\bibinfo{author}{\bibfnamefont{T.~P.} \bibnamefont{{Sotiriou}}}
  \bibnamefont{and}
  \bibinfo{author}{\bibfnamefont{V.}~\bibnamefont{{Faraoni}}},
  \bibinfo{journal}{Reviews of Modern Physics} \textbf{\bibinfo{volume}{82}},
  \bibinfo{pages}{451} (\bibinfo{year}{2010}), \eprint{0805.1726}.

\bibitem[{\citenamefont{{Nicolis} et~al.}(2009)\citenamefont{{Nicolis},
  {Rattazzi}, and {Trincherini}}}]{Galileon2009}
\bibinfo{author}{\bibfnamefont{A.}~\bibnamefont{{Nicolis}}},
  \bibinfo{author}{\bibfnamefont{R.}~\bibnamefont{{Rattazzi}}},
  \bibnamefont{and}
  \bibinfo{author}{\bibfnamefont{E.}~\bibnamefont{{Trincherini}}},
  \bibinfo{journal}{\prd} \textbf{\bibinfo{volume}{79}}, \bibinfo{eid}{064036}
  (\bibinfo{year}{2009}), \eprint{0811.2197}.

\bibitem[{\citenamefont{{Deffayet}
  et~al.}(2009{\natexlab{a}})\citenamefont{{Deffayet}, {Esposito-Far{\`e}se},
  and {Vikman}}}]{CovGal2009}
\bibinfo{author}{\bibfnamefont{C.}~\bibnamefont{{Deffayet}}},
  \bibinfo{author}{\bibfnamefont{G.}~\bibnamefont{{Esposito-Far{\`e}se}}},
  \bibnamefont{and} \bibinfo{author}{\bibfnamefont{A.}~\bibnamefont{{Vikman}}},
  \bibinfo{journal}{\prd} \textbf{\bibinfo{volume}{79}}, \bibinfo{eid}{084003}
  (\bibinfo{year}{2009}{\natexlab{a}}), \eprint{0901.1314}.

\bibitem[{\citenamefont{{Vainshtein}}(1972)}]{Vainshtein1972}
\bibinfo{author}{\bibfnamefont{A.~I.} \bibnamefont{{Vainshtein}}},
  \bibinfo{journal}{Physics Letters B} \textbf{\bibinfo{volume}{39}},
  \bibinfo{pages}{393} (\bibinfo{year}{1972}).

\bibitem[{\citenamefont{{Hu} and {Sawicki}}(2007)}]{hs2007}
\bibinfo{author}{\bibfnamefont{W.}~\bibnamefont{{Hu}}} \bibnamefont{and}
  \bibinfo{author}{\bibfnamefont{I.}~\bibnamefont{{Sawicki}}},
  \bibinfo{journal}{\prd} \textbf{\bibinfo{volume}{76}}, \bibinfo{eid}{064004}
  (\bibinfo{year}{2007}), \eprint{0705.1158}.

\bibitem[{\citenamefont{{Li} et~al.}(2013{\natexlab{b}})\citenamefont{{Li},
  {Barreira}, {Baugh}, {Hellwing}, {Koyama}, {Pascoli}, and
  {Zhao}}}]{Li2013_quartic}
\bibinfo{author}{\bibfnamefont{B.}~\bibnamefont{{Li}}},
  \bibinfo{author}{\bibfnamefont{A.}~\bibnamefont{{Barreira}}},
  \bibinfo{author}{\bibfnamefont{C.~M.} \bibnamefont{{Baugh}}},
  \bibinfo{author}{\bibfnamefont{W.~A.} \bibnamefont{{Hellwing}}},
  \bibinfo{author}{\bibfnamefont{K.}~\bibnamefont{{Koyama}}},
  \bibinfo{author}{\bibfnamefont{S.}~\bibnamefont{{Pascoli}}},
  \bibnamefont{and} \bibinfo{author}{\bibfnamefont{G.-B.}
  \bibnamefont{{Zhao}}}, \bibinfo{journal}{\jcap}
  \textbf{\bibinfo{volume}{11}}, \bibinfo{eid}{012}
  (\bibinfo{year}{2013}{\natexlab{b}}), \eprint{1308.3491}.

\bibitem[{\citenamefont{{Deffayet}
  et~al.}(2009{\natexlab{b}})\citenamefont{{Deffayet}, {Deser}, and
  {Esposito-Far{\`e}se}}}]{Deffayet:2009mn}
\bibinfo{author}{\bibfnamefont{C.}~\bibnamefont{{Deffayet}}},
  \bibinfo{author}{\bibfnamefont{S.}~\bibnamefont{{Deser}}}, \bibnamefont{and}
  \bibinfo{author}{\bibfnamefont{G.}~\bibnamefont{{Esposito-Far{\`e}se}}},
  \bibinfo{journal}{\prd} \textbf{\bibinfo{volume}{80}}, \bibinfo{eid}{064015}
  (\bibinfo{year}{2009}{\natexlab{b}}), \eprint{0906.1967}.

\bibitem[{\citenamefont{{Peebles}}(1980)}]{1980Peebles}
\bibinfo{author}{\bibfnamefont{P.~J.~E.} \bibnamefont{{Peebles}}},
  \emph{\bibinfo{title}{{The large-scale structure of the universe}}}
  (\bibinfo{publisher}{Research supported by the National Science
  Foundation.~Princeton, N.J., Princeton University Press, 1980.~435 p.},
  \bibinfo{year}{1980}).

\bibitem[{\citenamefont{{Davis} and {Peebles}}(1977)}]{DavisPeebles_BBGKY}
\bibinfo{author}{\bibfnamefont{M.}~\bibnamefont{{Davis}}} \bibnamefont{and}
  \bibinfo{author}{\bibfnamefont{P.~J.~E.} \bibnamefont{{Peebles}}},
  \bibinfo{journal}{\apjs} \textbf{\bibinfo{volume}{34}}, \bibinfo{pages}{425}
  (\bibinfo{year}{1977}).

\bibitem[{\citenamefont{{Bogoliubov}}(1946)}]{bogoliubov1946}
\bibinfo{author}{\bibfnamefont{N.~N.} \bibnamefont{{Bogoliubov}}},
  \bibinfo{journal}{J. Phys.(USSR)} \textbf{\bibinfo{volume}{10}},
  \bibinfo{pages}{265} (\bibinfo{year}{1946}).

\bibitem[{\citenamefont{{Born} and {Green}}(1946)}]{BornGreen1946}
\bibinfo{author}{\bibfnamefont{M.}~\bibnamefont{{Born}}} \bibnamefont{and}
  \bibinfo{author}{\bibfnamefont{H.~S.} \bibnamefont{{Green}}},
  \bibinfo{journal}{Proc. R. Soc. Lond. A} \textbf{\bibinfo{volume}{188}},
  \bibinfo{pages}{10} (\bibinfo{year}{1946}).

\bibitem[{\citenamefont{{Kirkwood}}(1946)}]{Kirkwood1946}
\bibinfo{author}{\bibfnamefont{J.~G.} \bibnamefont{{Kirkwood}}},
  \bibinfo{journal}{\jcp} \textbf{\bibinfo{volume}{14}}, \bibinfo{pages}{180}
  (\bibinfo{year}{1946}).

\bibitem[{\citenamefont{{Yvon}}(1935)}]{yvon1935}
\bibinfo{author}{\bibfnamefont{J.}~\bibnamefont{{Yvon}}},
  \emph{\bibinfo{title}{{La th{\'e}orie statistique des fluides et
  l'{\'e}quation d'{\'e}tat}}}, vol. \bibinfo{volume}{203}
  (\bibinfo{publisher}{Hermann \& cie}, \bibinfo{year}{1935}).

\bibitem[{\citenamefont{{Juszkiewicz} et~al.}(1999)\citenamefont{{Juszkiewicz},
  {Springel}, and {Durrer}}}]{Juszkiewicz1999}
\bibinfo{author}{\bibfnamefont{R.}~\bibnamefont{{Juszkiewicz}}},
  \bibinfo{author}{\bibfnamefont{V.}~\bibnamefont{{Springel}}},
  \bibnamefont{and} \bibinfo{author}{\bibfnamefont{R.}~\bibnamefont{{Durrer}}},
  \bibinfo{journal}{\apjl} \textbf{\bibinfo{volume}{518}}, \bibinfo{pages}{L25}
  (\bibinfo{year}{1999}), \eprint{astro-ph/9812387}.

\bibitem[{\citenamefont{{Jenkins} et~al.}(1998)\citenamefont{{Jenkins},
  {Frenk}, {Pearce}, {Thomas}, {Colberg}, {White}, {Couchman}, {Peacock},
  {Efstathiou}, and {Nelson}}}]{Jenkins}
\bibinfo{author}{\bibfnamefont{A.}~\bibnamefont{{Jenkins}}},
  \bibinfo{author}{\bibfnamefont{C.~S.} \bibnamefont{{Frenk}}},
  \bibinfo{author}{\bibfnamefont{F.~R.} \bibnamefont{{Pearce}}},
  \bibinfo{author}{\bibfnamefont{P.~A.} \bibnamefont{{Thomas}}},
  \bibinfo{author}{\bibfnamefont{J.~M.} \bibnamefont{{Colberg}}},
  \bibinfo{author}{\bibfnamefont{S.~D.~M.} \bibnamefont{{White}}},
  \bibinfo{author}{\bibfnamefont{H.~M.~P.} \bibnamefont{{Couchman}}},
  \bibinfo{author}{\bibfnamefont{J.~A.} \bibnamefont{{Peacock}}},
  \bibinfo{author}{\bibfnamefont{G.}~\bibnamefont{{Efstathiou}}},
  \bibnamefont{and} \bibinfo{author}{\bibfnamefont{A.~H.}
  \bibnamefont{{Nelson}}}, \bibinfo{journal}{\apj}
  \textbf{\bibinfo{volume}{499}}, \bibinfo{pages}{20} (\bibinfo{year}{1998}),
  \eprint{astro-ph/9709010}.

\bibitem[{\citenamefont{{Kravtsov} et~al.}(2004)\citenamefont{{Kravtsov},
  {Berlind}, {Wechsler}, {Klypin}, {Gottl{\"o}ber}, {Allgood}, and
  {Primack}}}]{Kravtsov2004}
\bibinfo{author}{\bibfnamefont{A.~V.} \bibnamefont{{Kravtsov}}},
  \bibinfo{author}{\bibfnamefont{A.~A.} \bibnamefont{{Berlind}}},
  \bibinfo{author}{\bibfnamefont{R.~H.} \bibnamefont{{Wechsler}}},
  \bibinfo{author}{\bibfnamefont{A.~A.} \bibnamefont{{Klypin}}},
  \bibinfo{author}{\bibfnamefont{S.}~\bibnamefont{{Gottl{\"o}ber}}},
  \bibinfo{author}{\bibfnamefont{B.}~\bibnamefont{{Allgood}}},
  \bibnamefont{and} \bibinfo{author}{\bibfnamefont{J.~R.}
  \bibnamefont{{Primack}}}, \bibinfo{journal}{\apj}
  \textbf{\bibinfo{volume}{609}}, \bibinfo{pages}{35} (\bibinfo{year}{2004}),
  \eprint{astro-ph/0308519}.

\bibitem[{\citenamefont{{Reid} et~al.}(2010)\citenamefont{{Reid}, {Percival},
  {Eisenstein}, {Verde}, {Spergel}, {Skibba}, {Bahcall}, {Budavari}, {Frieman},
  {Fukugita} et~al.}}]{SDSS_LRGs}
\bibinfo{author}{\bibfnamefont{B.~A.} \bibnamefont{{Reid}}},
  \bibinfo{author}{\bibfnamefont{W.~J.} \bibnamefont{{Percival}}},
  \bibinfo{author}{\bibfnamefont{D.~J.} \bibnamefont{{Eisenstein}}},
  \bibinfo{author}{\bibfnamefont{L.}~\bibnamefont{{Verde}}},
  \bibinfo{author}{\bibfnamefont{D.~N.} \bibnamefont{{Spergel}}},
  \bibinfo{author}{\bibfnamefont{R.~A.} \bibnamefont{{Skibba}}},
  \bibinfo{author}{\bibfnamefont{N.~A.} \bibnamefont{{Bahcall}}},
  \bibinfo{author}{\bibfnamefont{T.}~\bibnamefont{{Budavari}}},
  \bibinfo{author}{\bibfnamefont{J.~A.} \bibnamefont{{Frieman}}},
  \bibinfo{author}{\bibfnamefont{M.}~\bibnamefont{{Fukugita}}},
  \bibnamefont{et~al.}, \bibinfo{journal}{\mnras}
  \textbf{\bibinfo{volume}{404}}, \bibinfo{pages}{60} (\bibinfo{year}{2010}),
  \eprint{0907.1659}.

\bibitem[{\citenamefont{{Wake} et~al.}(2008)\citenamefont{{Wake}, {Sheth},
  {Nichol}, {Baugh}, {Bland-Hawthorn}, {Colless}, {Couch}, {Croom}, {de
  Propris}, {Drinkwater} et~al.}}]{Wake2008}
\bibinfo{author}{\bibfnamefont{D.~A.} \bibnamefont{{Wake}}},
  \bibinfo{author}{\bibfnamefont{R.~K.} \bibnamefont{{Sheth}}},
  \bibinfo{author}{\bibfnamefont{R.~C.} \bibnamefont{{Nichol}}},
  \bibinfo{author}{\bibfnamefont{C.~M.} \bibnamefont{{Baugh}}},
  \bibinfo{author}{\bibfnamefont{J.}~\bibnamefont{{Bland-Hawthorn}}},
  \bibinfo{author}{\bibfnamefont{M.}~\bibnamefont{{Colless}}},
  \bibinfo{author}{\bibfnamefont{W.~J.} \bibnamefont{{Couch}}},
  \bibinfo{author}{\bibfnamefont{S.~M.} \bibnamefont{{Croom}}},
  \bibinfo{author}{\bibfnamefont{R.}~\bibnamefont{{de Propris}}},
  \bibinfo{author}{\bibfnamefont{M.~J.} \bibnamefont{{Drinkwater}}},
  \bibnamefont{et~al.}, \bibinfo{journal}{\mnras}
  \textbf{\bibinfo{volume}{387}}, \bibinfo{pages}{1045} (\bibinfo{year}{2008}),
  \eprint{0802.4288}.

\bibitem[{\citenamefont{{Barreira} et~al.}(2014)\citenamefont{{Barreira}, {Li},
  {Hellwing}, {Lombriser}, {Baugh}, and {Pascoli}}}]{Barreira_hm_2014}
\bibinfo{author}{\bibfnamefont{A.}~\bibnamefont{{Barreira}}},
  \bibinfo{author}{\bibfnamefont{B.}~\bibnamefont{{Li}}},
  \bibinfo{author}{\bibfnamefont{W.~A.} \bibnamefont{{Hellwing}}},
  \bibinfo{author}{\bibfnamefont{L.}~\bibnamefont{{Lombriser}}},
  \bibinfo{author}{\bibfnamefont{C.~M.} \bibnamefont{{Baugh}}},
  \bibnamefont{and}
  \bibinfo{author}{\bibfnamefont{S.}~\bibnamefont{{Pascoli}}},
  \bibinfo{journal}{ArXiv e-prints}  (\bibinfo{year}{2014}),
  \eprint{1401.1497}.

\bibitem[{\citenamefont{{Scoccimarro}}(2004)}]{Scoccimarro2004}
\bibinfo{author}{\bibfnamefont{R.}~\bibnamefont{{Scoccimarro}}},
  \bibinfo{journal}{\prd} \textbf{\bibinfo{volume}{70}}, \bibinfo{eid}{083007}
  (\bibinfo{year}{2004}), \eprint{astro-ph/0407214}.

\bibitem[{\citenamefont{{Barreira}
  et~al.}(2013{\natexlab{b}})\citenamefont{{Barreira}, {Li}, {Baugh}, and
  {Pascoli}}}]{Barreira_sph_col_2013}
\bibinfo{author}{\bibfnamefont{A.}~\bibnamefont{{Barreira}}},
  \bibinfo{author}{\bibfnamefont{B.}~\bibnamefont{{Li}}},
  \bibinfo{author}{\bibfnamefont{C.~M.} \bibnamefont{{Baugh}}},
  \bibnamefont{and}
  \bibinfo{author}{\bibfnamefont{S.}~\bibnamefont{{Pascoli}}},
  \bibinfo{journal}{\jcap} \textbf{\bibinfo{volume}{11}}, \bibinfo{eid}{056}
  (\bibinfo{year}{2013}{\natexlab{b}}), \eprint{1308.3699}.

\bibitem[{\citenamefont{{Aihara} et~al.}(2011)\citenamefont{{Aihara}, {Allende
  Prieto}, {An}, {Anderson}, {Aubourg}, {Balbinot}, {Beers}, {Berlind},
  {Bickerton}, {Bizyaev} et~al.}}]{SDSSDR8}
\bibinfo{author}{\bibfnamefont{H.}~\bibnamefont{{Aihara}}},
  \bibinfo{author}{\bibfnamefont{C.}~\bibnamefont{{Allende Prieto}}},
  \bibinfo{author}{\bibfnamefont{D.}~\bibnamefont{{An}}},
  \bibinfo{author}{\bibfnamefont{S.~F.} \bibnamefont{{Anderson}}},
  \bibinfo{author}{\bibfnamefont{{\'E}.}~\bibnamefont{{Aubourg}}},
  \bibinfo{author}{\bibfnamefont{E.}~\bibnamefont{{Balbinot}}},
  \bibinfo{author}{\bibfnamefont{T.~C.} \bibnamefont{{Beers}}},
  \bibinfo{author}{\bibfnamefont{A.~A.} \bibnamefont{{Berlind}}},
  \bibinfo{author}{\bibfnamefont{S.~J.} \bibnamefont{{Bickerton}}},
  \bibinfo{author}{\bibfnamefont{D.}~\bibnamefont{{Bizyaev}}},
  \bibnamefont{et~al.}, \bibinfo{journal}{\apjs}
  \textbf{\bibinfo{volume}{193}}, \bibinfo{eid}{29} (\bibinfo{year}{2011}),
  \eprint{1101.1559}.

\bibitem[{\citenamefont{{Kazin} et~al.}(2013)\citenamefont{{Kazin},
  {S{\'a}nchez}, {Cuesta}, {Beutler}, {Chuang}, {Eisenstein}, {Manera},
  {Padmanabhan}, {Percival}, {Prada} et~al.}}]{BOSS1}
\bibinfo{author}{\bibfnamefont{E.~A.} \bibnamefont{{Kazin}}},
  \bibinfo{author}{\bibfnamefont{A.~G.} \bibnamefont{{S{\'a}nchez}}},
  \bibinfo{author}{\bibfnamefont{A.~J.} \bibnamefont{{Cuesta}}},
  \bibinfo{author}{\bibfnamefont{F.}~\bibnamefont{{Beutler}}},
  \bibinfo{author}{\bibfnamefont{C.-H.} \bibnamefont{{Chuang}}},
  \bibinfo{author}{\bibfnamefont{D.~J.} \bibnamefont{{Eisenstein}}},
  \bibinfo{author}{\bibfnamefont{M.}~\bibnamefont{{Manera}}},
  \bibinfo{author}{\bibfnamefont{N.}~\bibnamefont{{Padmanabhan}}},
  \bibinfo{author}{\bibfnamefont{W.~J.} \bibnamefont{{Percival}}},
  \bibinfo{author}{\bibfnamefont{F.}~\bibnamefont{{Prada}}},
  \bibnamefont{et~al.}, \bibinfo{journal}{\mnras}
  \textbf{\bibinfo{volume}{435}}, \bibinfo{pages}{64} (\bibinfo{year}{2013}),
  \eprint{1303.4391}.

\bibitem[{\citenamefont{{Anderson} et~al.}(2014)\citenamefont{{Anderson},
  {Aubourg}, {Bailey}, {Beutler}, {Bolton}, {Brinkmann}, {Brownstein},
  {Chuang}, {Cuesta}, {Dawson} et~al.}}]{BOSS2}
\bibinfo{author}{\bibfnamefont{L.}~\bibnamefont{{Anderson}}},
  \bibinfo{author}{\bibfnamefont{E.}~\bibnamefont{{Aubourg}}},
  \bibinfo{author}{\bibfnamefont{S.}~\bibnamefont{{Bailey}}},
  \bibinfo{author}{\bibfnamefont{F.}~\bibnamefont{{Beutler}}},
  \bibinfo{author}{\bibfnamefont{A.~S.} \bibnamefont{{Bolton}}},
  \bibinfo{author}{\bibfnamefont{J.}~\bibnamefont{{Brinkmann}}},
  \bibinfo{author}{\bibfnamefont{J.~R.} \bibnamefont{{Brownstein}}},
  \bibinfo{author}{\bibfnamefont{C.-H.} \bibnamefont{{Chuang}}},
  \bibinfo{author}{\bibfnamefont{A.~J.} \bibnamefont{{Cuesta}}},
  \bibinfo{author}{\bibfnamefont{K.~S.} \bibnamefont{{Dawson}}},
  \bibnamefont{et~al.}, \bibinfo{journal}{\mnras}
  \textbf{\bibinfo{volume}{439}}, \bibinfo{pages}{83} (\bibinfo{year}{2014}),
  \eprint{1303.4666}.

\bibitem[{\citenamefont{{Metcalfe} et~al.}(2013)\citenamefont{{Metcalfe},
  {Farrow}, {Cole}, {Draper}, {Norberg}, {Burgett}, {Chambers}, {Denneau},
  {Flewelling}, {Kaiser} et~al.}}]{PANSTARRS1}
\bibinfo{author}{\bibfnamefont{N.}~\bibnamefont{{Metcalfe}}},
  \bibinfo{author}{\bibfnamefont{D.~J.} \bibnamefont{{Farrow}}},
  \bibinfo{author}{\bibfnamefont{S.}~\bibnamefont{{Cole}}},
  \bibinfo{author}{\bibfnamefont{P.~W.} \bibnamefont{{Draper}}},
  \bibinfo{author}{\bibfnamefont{P.}~\bibnamefont{{Norberg}}},
  \bibinfo{author}{\bibfnamefont{W.~S.} \bibnamefont{{Burgett}}},
  \bibinfo{author}{\bibfnamefont{K.~C.} \bibnamefont{{Chambers}}},
  \bibinfo{author}{\bibfnamefont{L.}~\bibnamefont{{Denneau}}},
  \bibinfo{author}{\bibfnamefont{H.}~\bibnamefont{{Flewelling}}},
  \bibinfo{author}{\bibfnamefont{N.}~\bibnamefont{{Kaiser}}},
  \bibnamefont{et~al.}, \bibinfo{journal}{\mnras}
  \textbf{\bibinfo{volume}{435}}, \bibinfo{pages}{1825} (\bibinfo{year}{2013}),
  \eprint{1310.6368}.

\bibitem[{\citenamefont{{Farrow} et~al.}(2014)\citenamefont{{Farrow}, {Cole},
  {Metcalfe}, {Draper}, {Norberg}, {Foucaud}, {Burgett}, {Chambers}, {Kaiser},
  {Kudritzki} et~al.}}]{PANSTARRS2}
\bibinfo{author}{\bibfnamefont{D.~J.} \bibnamefont{{Farrow}}},
  \bibinfo{author}{\bibfnamefont{S.}~\bibnamefont{{Cole}}},
  \bibinfo{author}{\bibfnamefont{N.}~\bibnamefont{{Metcalfe}}},
  \bibinfo{author}{\bibfnamefont{P.~W.} \bibnamefont{{Draper}}},
  \bibinfo{author}{\bibfnamefont{P.}~\bibnamefont{{Norberg}}},
  \bibinfo{author}{\bibfnamefont{S.}~\bibnamefont{{Foucaud}}},
  \bibinfo{author}{\bibfnamefont{W.~S.} \bibnamefont{{Burgett}}},
  \bibinfo{author}{\bibfnamefont{K.~C.} \bibnamefont{{Chambers}}},
  \bibinfo{author}{\bibfnamefont{N.}~\bibnamefont{{Kaiser}}},
  \bibinfo{author}{\bibfnamefont{R.~P.} \bibnamefont{{Kudritzki}}},
  \bibnamefont{et~al.}, \bibinfo{journal}{\mnras}
  \textbf{\bibinfo{volume}{437}}, \bibinfo{pages}{748} (\bibinfo{year}{2014}),
  \eprint{1310.6366}.

\bibitem[{\citenamefont{{Davis} and {Peebles}}(1983)}]{1983Davis_Peebles}
\bibinfo{author}{\bibfnamefont{M.}~\bibnamefont{{Davis}}} \bibnamefont{and}
  \bibinfo{author}{\bibfnamefont{P.~J.~E.} \bibnamefont{{Peebles}}},
  \bibinfo{journal}{\apj} \textbf{\bibinfo{volume}{267}}, \bibinfo{pages}{465}
  (\bibinfo{year}{1983}).

\bibitem[{\citenamefont{{Benson} et~al.}(2000)\citenamefont{{Benson}, {Baugh},
  {Cole}, {Frenk}, and {Lacey}}}]{Benson2000_pvd_gal}
\bibinfo{author}{\bibfnamefont{A.~J.} \bibnamefont{{Benson}}},
  \bibinfo{author}{\bibfnamefont{C.~M.} \bibnamefont{{Baugh}}},
  \bibinfo{author}{\bibfnamefont{S.}~\bibnamefont{{Cole}}},
  \bibinfo{author}{\bibfnamefont{C.~S.} \bibnamefont{{Frenk}}},
  \bibnamefont{and} \bibinfo{author}{\bibfnamefont{C.~G.}
  \bibnamefont{{Lacey}}}, \bibinfo{journal}{\mnras}
  \textbf{\bibinfo{volume}{316}}, \bibinfo{pages}{107} (\bibinfo{year}{2000}),
  \eprint{astro-ph/9910488}.

\bibitem[{\citenamefont{{Li} et~al.}(2006)\citenamefont{{Li}, {Jing},
  {Kauffmann}, {B{\"o}rner}, {White}, and {Cheng}}}]{Li2006_pvd_gal}
\bibinfo{author}{\bibfnamefont{C.}~\bibnamefont{{Li}}},
  \bibinfo{author}{\bibfnamefont{Y.~P.} \bibnamefont{{Jing}}},
  \bibinfo{author}{\bibfnamefont{G.}~\bibnamefont{{Kauffmann}}},
  \bibinfo{author}{\bibfnamefont{G.}~\bibnamefont{{B{\"o}rner}}},
  \bibinfo{author}{\bibfnamefont{S.~D.~M.} \bibnamefont{{White}}},
  \bibnamefont{and} \bibinfo{author}{\bibfnamefont{F.~Z.}
  \bibnamefont{{Cheng}}}, \bibinfo{journal}{\mnras}
  \textbf{\bibinfo{volume}{368}}, \bibinfo{pages}{37} (\bibinfo{year}{2006}),
  \eprint{astro-ph/0509874}.

\bibitem[{\citenamefont{{Tinker} et~al.}(2007)\citenamefont{{Tinker},
  {Norberg}, {Weinberg}, and {Warren}}}]{Tinker2007_pvd_gal}
\bibinfo{author}{\bibfnamefont{J.~L.} \bibnamefont{{Tinker}}},
  \bibinfo{author}{\bibfnamefont{P.}~\bibnamefont{{Norberg}}},
  \bibinfo{author}{\bibfnamefont{D.~H.} \bibnamefont{{Weinberg}}},
  \bibnamefont{and} \bibinfo{author}{\bibfnamefont{M.~S.}
  \bibnamefont{{Warren}}}, \bibinfo{journal}{\apj}
  \textbf{\bibinfo{volume}{659}}, \bibinfo{pages}{877} (\bibinfo{year}{2007}),
  \eprint{astro-ph/0603543}.

\bibitem[{\citenamefont{{de la Torre} and {Guzzo}}(2012)}]{delaTorre_rsd}
\bibinfo{author}{\bibfnamefont{S.}~\bibnamefont{{de la Torre}}}
  \bibnamefont{and} \bibinfo{author}{\bibfnamefont{L.}~\bibnamefont{{Guzzo}}},
  \bibinfo{journal}{\mnras} \textbf{\bibinfo{volume}{427}},
  \bibinfo{pages}{327} (\bibinfo{year}{2012}), \eprint{1202.5559}.

\bibitem[{\citenamefont{{Strauss} and {Willick}}(1995)}]{StraussWillick}
\bibinfo{author}{\bibfnamefont{M.~A.} \bibnamefont{{Strauss}}}
  \bibnamefont{and} \bibinfo{author}{\bibfnamefont{J.~A.}
  \bibnamefont{{Willick}}}, \bibinfo{journal}{\physrep}
  \textbf{\bibinfo{volume}{261}}, \bibinfo{pages}{271} (\bibinfo{year}{1995}),
  \eprint{astro-ph/9502079}.

\bibitem[{\citenamefont{{Tully} et~al.}(2013)\citenamefont{{Tully}, {Courtois},
  {Dolphin}, {Fisher}, {H{\'e}raudeau}, {Jacobs}, {Karachentsev}, {Makarov},
  {Makarova}, {Mitronova} et~al.}}]{Tully2013}
\bibinfo{author}{\bibfnamefont{R.~B.} \bibnamefont{{Tully}}},
  \bibinfo{author}{\bibfnamefont{H.~M.} \bibnamefont{{Courtois}}},
  \bibinfo{author}{\bibfnamefont{A.~E.} \bibnamefont{{Dolphin}}},
  \bibinfo{author}{\bibfnamefont{J.~R.} \bibnamefont{{Fisher}}},
  \bibinfo{author}{\bibfnamefont{P.}~\bibnamefont{{H{\'e}raudeau}}},
  \bibinfo{author}{\bibfnamefont{B.~A.} \bibnamefont{{Jacobs}}},
  \bibinfo{author}{\bibfnamefont{I.~D.} \bibnamefont{{Karachentsev}}},
  \bibinfo{author}{\bibfnamefont{D.}~\bibnamefont{{Makarov}}},
  \bibinfo{author}{\bibfnamefont{L.}~\bibnamefont{{Makarova}}},
  \bibinfo{author}{\bibfnamefont{S.}~\bibnamefont{{Mitronova}}},
  \bibnamefont{et~al.}, \bibinfo{journal}{\aj} \textbf{\bibinfo{volume}{146}},
  \bibinfo{eid}{86} (\bibinfo{year}{2013}), \eprint{1307.7213}.

\bibitem[{\citenamefont{{Nusser} et~al.}(2012)\citenamefont{{Nusser},
  {Branchini}, and {Davis}}}]{Nusser2012}
\bibinfo{author}{\bibfnamefont{A.}~\bibnamefont{{Nusser}}},
  \bibinfo{author}{\bibfnamefont{E.}~\bibnamefont{{Branchini}}},
  \bibnamefont{and} \bibinfo{author}{\bibfnamefont{M.}~\bibnamefont{{Davis}}},
  \bibinfo{journal}{\apj} \textbf{\bibinfo{volume}{755}}, \bibinfo{eid}{58}
  (\bibinfo{year}{2012}), \eprint{1202.4138}.

\bibitem[{\citenamefont{{Gorski} et~al.}(1989)\citenamefont{{Gorski}, {Davis},
  {Strauss}, {White}, and {Yahil}}}]{Gorski1989}
\bibinfo{author}{\bibfnamefont{K.~M.} \bibnamefont{{Gorski}}},
  \bibinfo{author}{\bibfnamefont{M.}~\bibnamefont{{Davis}}},
  \bibinfo{author}{\bibfnamefont{M.~A.} \bibnamefont{{Strauss}}},
  \bibinfo{author}{\bibfnamefont{S.~D.~M.} \bibnamefont{{White}}},
  \bibnamefont{and} \bibinfo{author}{\bibfnamefont{A.}~\bibnamefont{{Yahil}}},
  \bibinfo{journal}{\apj} \textbf{\bibinfo{volume}{344}}, \bibinfo{pages}{1}
  (\bibinfo{year}{1989}).

\bibitem[{\citenamefont{{Juszkiewicz} et~al.}(2000)\citenamefont{{Juszkiewicz},
  {Ferreira}, {Feldman}, {Jaffe}, and {Davis}}}]{Juszkiewicz2000}
\bibinfo{author}{\bibfnamefont{R.}~\bibnamefont{{Juszkiewicz}}},
  \bibinfo{author}{\bibfnamefont{P.~G.} \bibnamefont{{Ferreira}}},
  \bibinfo{author}{\bibfnamefont{H.~A.} \bibnamefont{{Feldman}}},
  \bibinfo{author}{\bibfnamefont{A.~H.} \bibnamefont{{Jaffe}}},
  \bibnamefont{and} \bibinfo{author}{\bibfnamefont{M.}~\bibnamefont{{Davis}}},
  \bibinfo{journal}{Science} \textbf{\bibinfo{volume}{287}},
  \bibinfo{pages}{109} (\bibinfo{year}{2000}), \eprint{astro-ph/0001041}.

\bibitem[{\citenamefont{{Feldman} et~al.}(2003)\citenamefont{{Feldman},
  {Juszkiewicz}, {Ferreira}, {Davis}, {Gazta{\~n}aga}, {Fry}, {Jaffe},
  {Chambers}, {da Costa}, {Bernardi} et~al.}}]{streaming_vel_Omega}
\bibinfo{author}{\bibfnamefont{H.}~\bibnamefont{{Feldman}}},
  \bibinfo{author}{\bibfnamefont{R.}~\bibnamefont{{Juszkiewicz}}},
  \bibinfo{author}{\bibfnamefont{P.}~\bibnamefont{{Ferreira}}},
  \bibinfo{author}{\bibfnamefont{M.}~\bibnamefont{{Davis}}},
  \bibinfo{author}{\bibfnamefont{E.}~\bibnamefont{{Gazta{\~n}aga}}},
  \bibinfo{author}{\bibfnamefont{J.}~\bibnamefont{{Fry}}},
  \bibinfo{author}{\bibfnamefont{A.}~\bibnamefont{{Jaffe}}},
  \bibinfo{author}{\bibfnamefont{S.}~\bibnamefont{{Chambers}}},
  \bibinfo{author}{\bibfnamefont{L.}~\bibnamefont{{da Costa}}},
  \bibinfo{author}{\bibfnamefont{M.}~\bibnamefont{{Bernardi}}},
  \bibnamefont{et~al.}, \bibinfo{journal}{\apjl}
  \textbf{\bibinfo{volume}{596}}, \bibinfo{pages}{L131} (\bibinfo{year}{2003}),
  \eprint{astro-ph/0305078}.

\bibitem[{\citenamefont{{Ferreira} et~al.}(1999)\citenamefont{{Ferreira},
  {Juszkiewicz}, {Feldman}, {Davis}, and {Jaffe}}}]{Ferreira1999}
\bibinfo{author}{\bibfnamefont{P.~G.} \bibnamefont{{Ferreira}}},
  \bibinfo{author}{\bibfnamefont{R.}~\bibnamefont{{Juszkiewicz}}},
  \bibinfo{author}{\bibfnamefont{H.~A.} \bibnamefont{{Feldman}}},
  \bibinfo{author}{\bibfnamefont{M.}~\bibnamefont{{Davis}}}, \bibnamefont{and}
  \bibinfo{author}{\bibfnamefont{A.~H.} \bibnamefont{{Jaffe}}},
  \bibinfo{journal}{\apjl} \textbf{\bibinfo{volume}{515}}, \bibinfo{pages}{L1}
  (\bibinfo{year}{1999}), \eprint{astro-ph/9812456}.

\end{thebibliography}

\acknowledgements The authors are grateful to anonymous
referees who helped improve the scientific quality of this work.
We have greatly benefited from discussions with Adi
Nusser, Enzo Branchini, Maciej Bilicki, Marius Cautun, Micha\l{}
Chodorowski, Nick Kaiser and John Peacock.  Maciej Bilicki and Marius
Cautun are warmly acknowledged for a careful reading of the manuscript
and Lydia Heck for her technical support for our computations. This
work used the DiRAC Data Centric system at Durham University, operated
by the Institute for Computational Cosmology on behalf of the STFC
DiRAC HPC Facility (\url{www.dirac.ac.uk}). This equipment was funded
by BIS National E-infrastructure capital grant ST/K00042X/1, STFC
capital grant ST/H008519/1, and STFC DiRAC Operations grant
ST/K003267/1 and Durham University. DiRAC is part of the National
E-Infrastructure programme. This work was supported by the Science and
Technology Facilities Council [grant number ST/F001166/1]; Polish
National Science Center [grant number DEC-2011/01/D/ST9/01960]; ERC
Advanced Investigator grant, COSMIWAY [grant number GA 267291 ], and
FCT-Portugal [grant number SFRH/BD/75791/2011]. This
research was carried out with the support of the “HPC
Infrastructure for Grand Challenges of Science and
Engineering” Project, co-financed by the European
Regional Development Fund under the Innovative
Economy Operational Programme.

\end{document}